\documentstyle[aps,pra,preprint]{revtex}

\begin{document}

\draft

\title{Excitation spectroscopy of vortex states in dilute 
Bose-condensed gases}

\author{R.\ J.\ Dodd\thanks{Also at Physics Laboratory, 
National Institute of Standards and Technology,
Technology Administration, 
U.S.\ Department of Commerce, Gaithersburg,
MD 20899.}}
\address{Institute for Physical Science and Technology,
University of Maryland at College Park,\\ 
College Park, MD 20742.}

\author{K.\ Burnett$^\ast$}
\address{Clarendon Laboratory, Department of Physics,
University of Oxford,\\ 
Parks Road, Oxford OX1 3PU, United Kingdom.}

\author{Mark Edwards$^\ast$}
\address{Department of Physics,
Georgia Southern University, Statesboro, GA 30460-8031.}

\author{Charles\ W.\ Clark}
\address{Electron and Optical Physics Division,
National Institute of Standards and Technology,\\ 
Technology Administration, U. S. Department of Commerce, 
Gaithersburg, MD 20899.}

\date{\today}

\maketitle

\begin{abstract}
We apply linear-response analysis to the Gross-Pitaevskii equation to
obtain the excitation frequencies of a Bose-Einstein condensate in a
vortex state, and apply it to a system of Rubidium atoms confined in a
time-averaged orbiting potential trap.  The excitation frequencies of
a vortex differ significantly from those of the ground state, and may
therefore be used to obtain a spectroscopic signature of the presence
of a vortex state.
\end{abstract}


%%%%%%%%%%%%%%%%%%%%%%%%%%%%%%%%%%%%%%%%%%%%%%%%%%%%%%%%%%%%%%%%%%%%%%
%                    		 Introduction              	     %
%%%%%%%%%%%%%%%%%%%%%%%%%%%%%%%%%%%%%%%%%%%%%%%%%%%%%%%%%%%%%%%%%%%%%%
The recent attainment of quantum degeneracy conditions in
magnetically-trapped, dilute alkali vapors
\cite{R_JILA_BEC1995,R_RICE_BEC1995,R_MIT_BEC1995,R_MIT2_BEC1996} has
opened an avenue for studying the many-body physics of Bose-Einstein
condensates (BECs) in unprecedented detail.  Recent experiments have
mapped out many of the basic properties of alkali BECs: the critical
temperature $T_c$ \cite{R_MIT2_BEC1996}, the temperature dependence of
the condensate fraction $N_0/N$ \cite{R_MIT2_BEC1996}, the
contribution of particle interactions to the internal energy (i.e.
departures from ideal-gas behavior) \cite{R_MIT2_BEC1996}, and the
energies of low-lying excitations \cite{R_JIN1996,R_MEWES1996}.

There remain many aspects of ``classical'' superfluid behavior that
have not yet been encountered in the atomic BECs.  The one we discuss
in this paper is the generation of vortices, which to our knowledge
have not yet been observed in trapped gases.  Recent theoretical
investigations of axially-symmetric harmonic traps have identified
vortex states of condensates which have sharp values of the azimuthal
component $m$ of the angular momentum
\cite{R_EDWARDS1996,R_DALFOVO1996,R_DODD_NISTJR_1996}.  Rotation of
the trap at a critical frequency $\omega_{\rm crit}$, which is of the
order of the harmonic trap frequency $\omega$, should force the
condensate into the vortex state
\cite{R_EDWARDS1996,R_DALFOVO1996,R_DODD_NISTJR_1996}.  Detection of
this state by current imaging techniques is complicated by present
magnet geometries, which constrain viewing of a condensate to be done
more or less perpendicular to the trap symmetry axis (hereafter the
$z$ axis).  Most schemes acquire images of the condensate density
integrated along the line of sight, and the integrated density of a
vortex state perpendicular to the trap axis is not much different from
that of the condensate ground state.  In this paper we calculate the
mechanical excitation spectrum of a vortex state, and find that it
differs significantly from that of the ground state.  We thus propose
excitation spectroscopy as a sensitive technique for detecting the
presence of vortices.

%%%%%%%%%%%%%%%%%%%%%%%%%%%%%%%%%%%%%%%%%%%%%%%%%%%%%%%%%%%%%%%%%%%%%%
%                   		 Method 	                     %
%%%%%%%%%%%%%%%%%%%%%%%%%%%%%%%%%%%%%%%%%%%%%%%%%%%%%%%%%%%%%%%%%%%%%%

The basic framework of our method is mean-field theory as described by
the Gross-Pitaevski (GP) formalism
\cite{R_GINZBURG1958,R_LIFSHITZ1980} for a condensate of a dilute Bose
gas at temperature $T = 0$.  Current experimental BEC atomic physics
appears to be practiced in a regime where this formalism is
applicable: the gases are very tenuous, and a nearly pure condensate
(corresponding to $T$ very near zero) can be produced by the technique
of evaporative cooling
\cite{R_JILA_BEC1995,R_MIT_BEC1995,R_MIT2_BEC1996}.  Calculations done
within the GP framework yield good agreement with experiment
concerning condensate shapes and sizes
\cite{R_EDWARDS1996,R_DALFOVO1996,R_HOLLAND1996,R_BAYM1996}, and give
ground-state condensate excitation frequencies within about 5\% of
experimental values
\cite{R_JIN1996,R_MEWES1996,R_EXC_EDWARDS1996}.

We begin with the time-independent treatment of condensate
eigenfunctions using the Gross-Pitaevskii (nonlinear Schr\"odinger)
equation, and then calculate the excitation spectrum using the method
of Bogoliubov \cite{R_BOGGLEYOURMIND1947}, which was used by
Pitaevskii to examine excitations about vortices in a homogeneous gas
\cite{R_PITAEVSKII1961}, and has been adapted to treat trapped Bose
condensates
\cite{R_EXC_EDWARDS1996,R_EXCITATION1996,R_FETTER1996,R_RUPRECHT1996}.
The specific calculations reported here are done for a system of
$^{87}$Rb atoms confined in the time-averaged orbiting potential (TOP)
trap currently in use at JILA \cite{R_PETRICH1995}.

In the GP formalism the interaction between atoms is approximated by
the pseudopotential, $V({\bf r},{\bf r}') = U_0 \delta({\bf r}-{\bf
r}')$, where $U_0 = 4 \pi \hbar^2 a / M$, with $M$ being the atomic
mass and $a$ the scattering length that characterizes low-energy
atomic collisions.  For the triplet Born-Oppenheimer potential of
$^{87}$Rb$_2$ that describes collisions in the JILA trap, the current
best estimate \cite{R_HEINZEN1996} of $a$ is $110 a_0$, where $a_0$ is
the Bohr radius.  This value is used in the present paper.  The
trapping potential takes the form $V_{\rm trap}({\bf r}) =
M(\omega_{\rho}^{2} \rho^2 + \omega_{z}^{2} z^2)/2$, in the
cylindrical coordinates appropriate to the TOP trap.  For this trap,
the axial and radial frequencies are always in the same ratio,
$\omega_{z}/\omega_{\rho} = \sqrt{8}$; it is convenient to
characterize the trap by the single parameter $\nu_{\rho} =
\omega_{\rho}/2\pi$, the value of which is often stated explicitly in
experimental papers.  In this paper we use $\nu_{\rho}$ = 74 Hz.

The time-independent GP equation that describes the condensate
wavefunction $\psi({\bf r})$ thus takes the form
\begin{equation}
\left( -\frac{\hbar^2}{2 M}\nabla^{2} + V_{\rm trap}({\bf r}) + N_0
U_{0} |\psi ({\bf r})|^{2} \right) \psi ({\bf r}) = \mu\psi({\bf r}),
\label{ground_state_nlse}
\end{equation}
where $N_0$ is the
average number of condensate atoms, and $\mu$ is the
chemical potential, which is treated as an eigenvalue;
the normalization condition
\begin{equation}
\int d{\bf r} \left| \psi({\bf r}) \right|^2 = 1
\label{GP_normalization}
\end{equation}
is implied.  The form of Eq. (\ref{ground_state_nlse}) is consistent
with the existence of solutions
\begin{equation}
\psi({\bf r}) = \psi^{(m)}(\rho,z) \frac{e^{im\phi}}{\sqrt{2\pi}}
\label{m_eigenfunction}
\end{equation}
that are eigenfunctions (with eigenvalues $\hbar m$) of the
azimuthal angular momentum operator $l_z$.
Previous work on this system 
\cite{R_EDWARDS1996,R_DALFOVO1996,R_DODD_NISTJR_1996,%
R_HOLLAND1996,R_BAYM1996}
has identified a solution with
$m = 0$ as the condensate ground state, and those with
$\left| m  \right| = 1$ as the lowest vortex solutions.

As discussed elsewhere \cite{R_BIGNIST1996}, we solve Eq.\
(\ref{ground_state_nlse}) by ``growing'' a condensate up from the
noninteracting case ($U_0 = 0$), in which the solution is given by
harmonic oscillator wave functions. The growth process may be
visualized as a gradual turning-on of the interaction strength $U_0$
or as a gradual increase in the number of condensate atoms $N_0$; the
two pictures are equivalent for this purpose since $N_0$ and $U_0$
appear in the equations of motion only through the product $N_0U_0$.
In the noninteracting case, $m$ is a good quantum number, and it is
preserved during the growth process.  We start with
$\psi^{(m)}(\rho,z)$ as an eigenfunction of the axially symmetric
three-dimensional harmonic oscillator with quantum numbers $m$, the
number $n_z$ of nodes in $z$, and the number $n_{\rho}$ of nodes in
the cylindrical radial coordinate $\rho$.  The lowest vortex state
with $m = 1$ has $n_z = n_{\rho} = 0$; the number of nodes in $\rho$
and $z$ is also found to be conserved during the growth process.  Our
general representation of $\psi^{(m)}(\rho,z)$ is as a linear
combination of oscillator eigenfunctions, and a set of nonlinear
equations is solved iteratively to determine the coefficients at each
value of $N_0 U_0$.  A cross-section of the density of the vortex
solution is shown in Fig. \ref{3DplotVortexState}.
  
With the solution of the time-independent GP equation in hand, we
calculate the response of the condensate to a weak oscillatory
perturbation by standard linear-response theory \cite{R_DEGENNES1966}.
The associated time-dependent driven GP equation takes the form:
\begin{eqnarray}
i\hbar\frac{\partial\Psi}{\partial t} &=& \Big[ H_{0} +
U_{0}|\Psi({\bf r},t)|^{2} \nonumber \\ &+& f({\bf r})e^{-i\omega_{p}
t} + f^*({\bf r})e^{i\omega_{p} t} \Big]
\Psi({\bf r},t),
\label{driven_ground_state_nlse}
\end{eqnarray}
where $f({\bf r})$ is the spatially-dependent amplitude of the
perturbation.  We solve this equation in the linear-response limit.
The details of this approach are described elsewhere
\cite{R_RUPRECHT1996}, and we simply state the central results here.
By using the form
\begin{equation}
\Psi({\bf r},t) = e^{-i\mu t/\hbar}
\left[
N_{0}^{\frac{1}{2}} \psi({\bf r}) + u({\bf r}) e^{-i\omega_{p} t} +
v^\ast({\bf r}) e^{i\omega_{p} t} \right]
\label{initial_form_of_wavefunction}
\end{equation}
we obtain Eq.\ (\ref{ground_state_nlse}) and also the linear-response
equations,
\begin{equation}
\left( {\cal L} - \hbar\omega_{p} \right) u({\bf r}) +
N_{0}U_{0}\left[\psi({\bf r})\right]^{2} v({\bf r}) = - N_0^{1/2}
f({\bf r})
\psi({\bf r}),
\label{u_equation}
\end{equation}
\begin{equation}
N_{0}U_{0}\left[\psi^{\ast}({\bf r})\right]^{2} u({\bf r}) +
\left( {\cal L} + \hbar\omega_{p} \right) v({\bf r}) =
\!- N_0^{1/2} f({\bf r})
\psi^\ast({\bf r}),
\label{v_equation}
\end{equation}
where ${\cal L} = H_{0} - \mu + 2 N_{0} U_0 \left|\psi({\bf
r})\right|^{2}$.

This pair of equations can be solved in a general way by expanding the
condensate response in normal modes of oscillation, which are obtained
by solving the Bogoliubov equations,
\begin{equation}
\left( {\cal L} - \hbar\omega_{\lambda} \right)
u_{\lambda}({\bf r}) + N_{0}U_{0}\left[\psi({\bf r})\right]
^{2}v_{\lambda}({\bf r}) = 0,
\label{eigenmode_u_equation}
\end{equation}
\begin{equation}
N_{0}U_{0}\left[\psi^{\ast}({\bf r})\right]^{2} u_{\lambda}({\bf r}) +
\left( {\cal L} + \hbar\omega_{\lambda}\right)
v_{\lambda}({\bf r}) = 0,
\label{eigenmode_v_equation}
\end{equation}
where $\omega_{\lambda}$ is an eigenvalue and $u_{\lambda}({\bf r})$,
$v_{\lambda}({\bf r})$ are corresponding eigenfunctions.  We require
that $u_{\lambda}({\bf r})$ and
$v_{\lambda}({\bf r})$ be square integrable and satisfy the
conventional normalization condition
\begin{equation}
\int d{\bf r}
\left( \left|u_{\lambda}({\bf r})\right|^{2} -
\left|v_{\lambda}({\bf r})\right|^{2} \right) = 1.
\label{norm_condition}
\end{equation}
With this condition in force, the condensate response to an arbitrary
driver $f({\bf r})$ is obtained by a superposition of normal
modes \cite{R_RUPRECHT1996},
\begin{equation}
{u({\bf r}) \choose v({\bf r})} = 
- \sum_\lambda 
\frac{g_\lambda}{\hbar \left( \omega_\lambda - w_p \right)}
{u_\lambda({\bf r}) 
\choose v_\lambda({\bf r})},
\label{drive_me}
\end{equation}
where the amplitudes $g_\lambda$ are obtained by quadrature: 
\begin{equation}
g_\lambda = N_0^{1/2} \int d{\bf r} 
f({\bf r}) \left[  \psi({\bf r})
u_\lambda^\ast({\bf r}) + \psi^\ast({\bf r})
v_\lambda^\ast({\bf r}) \right].
\label{selection_rules}
\end{equation}

Two key qualitative aspects of condensate excitation are implied by
Eqs.\ (\ref{drive_me}) and (\ref{selection_rules}).  First, from Eq.\
(\ref{drive_me}), the response is largest when the driving frequency
$\omega_p$ is equal to a normal-mode frequency $\omega_\lambda$ (the
apparent divergence in this response, as in the case of a driven
classical oscillator, is due to the neglect of damping in this
treatment).  Second, Eq.\ (\ref{selection_rules}) implies selection
rules for a given driver $f({\bf r})$, associated with symmetries of
the solutions of Eqs.\ (\ref{eigenmode_u_equation}) and
(\ref{eigenmode_v_equation}).  In particular, if $\psi({\bf r})$ is
given by Eq.\ (\ref{m_eigenfunction}), then it is straightforward to
show that the normal modes will have specific angular momentum
composition, in the following sense: if $u_\lambda({\bf r})$ is
an eigenfunction of $l_z$ with a particular eigenvalue $m_u$ (in units
of $\hbar$), then $v_\lambda({\bf r})$ will be an eigenfunction
with eigenvalue $m_u - 2 m$.  It is appropriate to think of a normal
mode as a quasiparticle moving in an effective potential created by
the condensate, and for $v_\lambda$ as being created by scattering of
$u_\lambda$ by the condensate.  For $m = 0$, the condensate has axial
symmetry, and the normal modes thus have definite values $m_u$ of the
angular momentum; for $m \neq 0$, the condensate is not axially
symmetric, and the component $u_\lambda$ (with angular momentum $m_u$)
is scattered into $v_\lambda$ (with angular momentum $m_u - 2 m$).  In
the case treated in this paper, we have $m = 1$, and we will label the
normal modes by the value of $m_u$ that corresponds to the component
$u_\lambda$.  Thus, a ``breathing-mode'' driver, such as was applied
to the ground state of the condensate in Ref. \cite{R_JIN1996}, will
result here in excitation of $m_u = 1$; a dipole driver, upon which we
comment below, yields $m_u = 0$ and 2; and a quadrupole driver, also
used in Ref. \cite{R_JIN1996}, gives $m_u = -1$ and 3.

We have solved Eqs.\ (\ref{eigenmode_u_equation} ) and
(\ref{eigenmode_v_equation}) by an extension of the technique used in
Ref. \cite{R_EXC_EDWARDS1996}, in which $u_\lambda$ and $v_\lambda$
are expanded in trap eigenfunctions of appropriate symmetry, to obtain
a system of linear eigenvalue equations.  Fig.\
\ref{3DplotVortexState} depicts $\psi$ along with $u_\lambda$ and
$v_\lambda$ as computed for the lowest normal mode, which has $m_u =
2$, and would therefore be generated from the vortex condensate by a
dipole excitation (e.g. oscillatory displacement of the center of the
trap).  The $s$- and $d$-wave characteristics of $v_\lambda$ and
$u_\lambda$ respectively are apparent in the figure.  An important
property of this normal mode, which differs from all cases we have
encountered in excitation of $m = 0$, is that its frequency
$\omega_\lambda$ is {\em less} than the trap frequency $\omega_\rho$.
The eigenfrequency of the lowest dipole mode for $m = 0$, on the other
hand, is identical to the trap frequency; that mode describes the
rigid oscillation of the condensate ground state.

The dependence of normal mode frequencies upon $N_0$ is depicted in
Fig.\ \ref{vortex_excitations}.  It shows two candidates for
excitation by dipole or quadrupole driving that have distinctive
spectral signatures, in that no excitation frequencies of the $m = 0$
condensate are nearby.  In addition to the $m_u = 2$ mode just
described, there is the $m_u = -1$ mode, which could be excited by a
quadrupole rotation in the sense {\em opposite} to that of the vortex.
This frequency vanishes in the noninteracting limit, due to the $m$
degeneracy of the cylindrical harmonic oscillator; its non-zero value
elsewhere results directly from interactions of the quasiparticle with
the condensate.

It is worth noting that the frequency spectrum in Fig.\ %
\ref{vortex_excitations} is applicable to a wide range of
TOP-trap geometries.  Solutions of Eqs.\ %
(\ref{eigenmode_u_equation}) and (\ref{eigenmode_v_equation}),
subject to the constraint of Eq.\ (\ref{norm_condition}),
apply to all TOP-trap geometries for which the parameter
\begin{equation}
\gamma = N_{0}a\left(M \nu_{\rho}\right)^{1/2}
\end{equation}
remains invariant~\cite{R_EXC_EDWARDS1996}.  This scaling 
law makes excitation spectroscopy a particularly effective 
tool for diagnosing the presence of vortex condensates.

In conclusion, we have calculated the excitation spectrum of vortex
states of dilute Bose-Einstein condensates, in the linear response
regime.  These spectra are sufficiently different from those for
ground state excitation, that they may provide useful diagnostics of
the presence of vortices in trapped atom systems.

%%%%%%%%%%%%%%%%%%%%%%%%%%%%%%%%%%%%%%%%%%%%%%%%%%%%%%%%%%%%%%%%%%%%%
%                           Acknowledgements                        %
%%%%%%%%%%%%%%%%%%%%%%%%%%%%%%%%%%%%%%%%%%%%%%%%%%%%%%%%%%%%%%%%%%%%%
\acknowledgements
We would like to thank E.\ A.\ Cornell and D.\ S.\ Jin for
enlightening conversations. This work was partly supported by National
Science Foundation grants PHY-9601261 and PHY-9612728, and by the
United Kingdom Engineering and Physical Sciences Research Council.

%%%%%%%%%%%%%%%%%%%%%%%%%%%%%%%%%%%%%%%%%%%%%%%%%%%%%%%%%%%%%%%%%%%%%%
%               	    References         	          	     %
%%%%%%%%%%%%%%%%%%%%%%%%%%%%%%%%%%%%%%%%%%%%%%%%%%%%%%%%%%%%%%%%%%%%%%

%%%%%%%%%%%%%%%%%%%%%%%%%%%%%%%%%%%%%%%%%%%%%%%%%%%%%%%%%%%%%%%%%%%%%%
%                   		Figures   		             %
%%%%%%%%%%%%%%%%%%%%%%%%%%%%%%%%%%%%%%%%%%%%%%%%%%%%%%%%%%%%%%%%%%%%%%
% Vortex Figure: XaxisCut_Vortex.eps
\begin{figure}
\caption{
The dashed line is a plot of $100 \left| v_\lambda \left( {\bf r}
\right) \right|^2$ for the lowest energy excitation, the dotted line
shows $100 \left| u_\lambda \left( {\bf r} \right) \right|^2$, the
factor of 100 is an arbitrary scaling factor used for plotting 
convenience. The solid line shows a plot of the spatial distribution 
of the $k = 1$ vortex number density, $n_0 \left( {\bf r} \right)$. 
In this plot $N_0 = 4939$ atoms, 
and $\nu_\rho = 74$ Hz for a TOP trap.}
\label{3DplotVortexState}
\end{figure}

% Vortex Excitations Frequencies Figure : excitations_vortex.eps
\begin{figure}
\caption{Normal-mode excitation frequencies $\omega_\lambda$ of the
vortex state in the TOP trap vs. number of $^{87}$Rb condensate atoms
$N_0$, in units of the trap frequency $\omega_\rho$.  Labels indicate
the angular momentum quantum number $m_u$ of the $u_\lambda$ component
of the normal-mode eigenfunction, as described in text.  There are two
degenerate frequencies at $\omega_\lambda = \omega_\rho$ for all
values of $N_0$, which correspond to rigid oscillations of the vortex
in the trap.}
\label{vortex_excitations}
\end{figure}

\end{document}